\documentclass[%
 reprint,
preprintnumbers,
 amsmath,amssymb,
 aps,
prb,
]{revtex4-2}
\usepackage{graphicx}
\usepackage{dcolumn}
\usepackage{bm}
\usepackage{hyperref}
\usepackage{amsmath,amssymb}
\hypersetup{colorlinks=true, linkcolor=blue, citecolor=blue, filecolor=blue,	urlcolor=blue}
\usepackage[normalem]{ulem}
\begin{document}

\title{Strongly correlated topological surface states in type-II Dirac semimetal NiTe$_{2}$}
\author{Neeraj Bhatt}
\affiliation{Department of Physics, Indian Institute of Science Education and Research Bhopal, Bhopal Bypass Road, Bhauri, Bhopal 462 066, India}%

\author{Asif Ali}
\affiliation{Department of Physics, Indian Institute of Science Education and Research Bhopal, Bhopal Bypass Road, Bhauri, Bhopal 462 066, India}%

\author{Deepali Sharma}
\affiliation{Department of Physics, Indian Institute of Science Education and Research Bhopal, Bhopal Bypass Road, Bhauri, Bhopal 462 066, India}%

\author{Sakshi Bansal}
\affiliation{Department of Physics, Indian Institute of Science Education and Research Bhopal, Bhopal Bypass Road, Bhauri, Bhopal 462 066, India}%

\author{Manasi Mandal}
\affiliation{Department of Physics, Indian Institute of Science Education and Research Bhopal, Bhopal Bypass Road, Bhauri, Bhopal 462 066, India}%

\author{Ravi Prakash Singh}
\affiliation{Department of Physics, Indian Institute of Science Education and Research Bhopal, Bhopal Bypass Road, Bhauri, Bhopal 462 066, India}%

\author{Ravi Shankar Singh}
\email{rssingh@iiserb.ac.in}
\affiliation{Department of Physics, Indian Institute of Science Education and Research Bhopal, Bhopal Bypass Road, Bhauri, Bhopal 462 066, India}%

\date{\today}

\begin{abstract}
Nontrivial topology in type-II Dirac semimetal NiTe$_2$ leading to topologically protected surface states give rise to fascinating phenomena holding great promise for next-generation electronic and spintronic devices. Key parameters $-$ such as lattice parameter, disorder, vacancies, and electron correlation $-$ significantly influence the electronic structure and, subsequently, the physical properties. To resolve the discrepancy between the theoretical description and experimentally observed topological surface states, we comprehensively investigate the electronic structure of NiTe$_2$ using angle-resolved photoemission spectroscopy and density functional theory. Although the bulk electronic structure is found to be well-described within mean field approaches, an accurate description of topological surface states is obtained only by incorporating surface electronic correlation. We reveal that the strongly correlated surface states forming  Dirac-like conical crossing much below Fermi level have hybridized Ni 3$d$ and Te 5$p$ character. These findings underscore the intricate interplay between electron correlation and band topology, broadening our understanding of many-body correlation effects on the topological surface states in quantum materials.
\end{abstract}

\maketitle
Topological insulators have metallic surface states, with a bulk bandgap appearing due to spin-orbit coupling (SOC) induced band inversion \cite{Zhang2009,RevModPhys.82.3045,RevModPhys.83.1057}. In contrast, Weyl semimetals (WSMs) have SOC-gapped bulk bands with isolated pairs of Weyl points, where the Berry curvature becomes singular \cite{PhysRevLett.107.127205,PhysRevX.5.031013,Paul2023}. WSMs require breaking time-reversal or lattice-inversion symmetry \cite{PhysRevB.85.165110,RevModPhys.88.021004}. Dirac semimetals (DSMs) can form when both symmetries are present, leading to degenerate Weyl points \cite{RevModPhys.90.015001,PhysRevB.85.195320,PhysRevLett.119.026404}. Type-I DSMs have a closed Fermi surface (FS) with vanishing density of states (DOS) at the Dirac point, while type-II DSMs have tilted Dirac cones with electron and hole-like FSs both contributing at the Dirac point, resulting in a finite DOS at the band crossing \cite{PhysRevB.96.041201,PhysRevLett.119.026404,PhysRevB.94.121117,https://doi.org/10.1002/qute.202400175}. 

Transition metal chalcogenides are at the forefront of the exploration of topological electronic structures and are predicted to have a striking influence of electron correlations on the band structure and topological properties \cite{PhysRevB.81.035122,PhysRevB.99.035123,PhysRevLett.121.136401,PhysRevB.110.125119}. Owing to potential significance, many-body interactions have also been explored in the trigonally-structured type-II Dirac semimetals 1T-$MX_2$ ($M$ = Ni, Pd, Pt; $X$ =  Te, Se) \cite{PhysRevB.105.115115,2023PhRvB.108w5138B,PhysRevB.106.075153,doi:10.1126/sciadv.abi6339,PhysRevB.110.165146,PhysRevLett.131.086501}.
 A bulk type-II Dirac point lies at $\sim$0.54 eV, $\sim$0.86 eV and $\sim$1.48 eV in PdTe$_2$, PtTe$_2$ and PtSe$_2$, respectively, below the Fermi level ($E_F$) \cite{PhysRevB.96.125102,PhysRevLett.119.016401,Yan2017,Bahramy2018,PhysRevLett.120.156401,PhysRevLett.120.156401}, while NiTe$_{2}$ hosts the Dirac point in close proximity to $E_F$ \cite{PhysRevB.100.195134,Xu2018,Mukherjee2020,PhysRevB.102.125103}. Apart from the type-II Dirac semimetallic phase, various topological surface states have been extensively investigated in these systems. Notably, two prominent surface states have been the subject of extensive experimental and theoretical discussion: one near $E_F$ along the $\overline{\Gamma}$-$\overline{\textrm{M}}$ direction (SS0), and another forming a Dirac-like conical crossing at $\overline{\Gamma}$ point (SS2) much below $E_F$ \cite{PhysRevB.100.195134,Yan2017,PhysRevLett.119.016401,Bahramy2018,Xu2018,Mukherjee2020,PhysRevLett.120.156401}. While, the bulk electronic structure has been argued to be described well within density functional theory (DFT), there appears to be a distinct deviation in the observed and theoretically predicted topological surface states. 
 For example, surface band structure using Green's function method based on tight-binding Hamiltonian mapped to DFT bands, exhibits large deviation in the energy position of SS2 in all the systems \cite{Bahramy2018,PhysRevB.100.195134,Xu2018,PhysRevLett.120.156401,PhysRevLett.120.156401}. Surface states calculated using Green's function approach is also found to be insensitive to the effect of in-plain strain in NiTe$_2$ \cite{PhysRevB.110.L201401}. 
  Though, surface band structure calculations within DFT, using slab configuration, provides better agreement with the experimental energy position of SS2, it is limited to Pd and Pt-based systems only \cite{doi:10.1021/acs.nanolett.2c04511}. For NiTe$_2$, Dirac crossing formed by SS2 has been consistently observed at $\sim$1.42 eV below $E_F$ (for samples grown under varied conditions), while the estimated position differs by $\sim$ 100 meV \cite{PhysRevB.104.155133,Mukherjee2020,10.1063/5.0016745}.

 The interplay of correlation effects and band topology stands out as a prominent research domains, exploring the emergent physics arising from their coexistence in transition metal complexes \cite{physRevB.83.205101,physRevLett.109.066401,NaturePhysics.16.641(2020),PhysRevLett.102.256403}.
For example, strong correlation induced alteration of band topology makes iron-based superconductors topologically nontrivial, contrary to initial predictions \cite{PhysRevB.106.115114}. Effect of GW-level many-body correlations on the band structure and topology of 1T-$MX_2$ results in a trivial metallic phase for PtSe$_2$ and PtTe$_2$ while type-II Dirac cone with substantial tilting is preserved in NiTe$_2$ \cite{PhysRevB.110.165146}. These observations along with energy position mismatch in surface states indicate electron correlation effects may be playing a role in NiTe$_2$ and are insignificant in 4$d$ and 5$d$ based transition metal dichalcogenides (TMDCs).

In this Letter, we systematically investigate the electronic structure of NiTe$_2$ using photoemission spectroscopy and DFT. Core levels and valence band spectra reveal the weakly-correlated semimetallic nature of NiTe$_2$. Angle-resolved photoemission spectroscopy (ARPES) reveal three-fold symmetric bulk FS along with various topological surface states. We unveil that the enhanced electronic correlation at the surface causes the observed discrepancy in SS2 and the energy position of the surface Dirac crossing could be efficiently captured within DFT+$U$ (on site Coulomb interaction) demonstrating the importance of electronic correlation effects in NiTe$_{2}$.
 
Photoemission spectroscopic measurements were performed on \textit{in-situ} cleaved NiTe$_{2}$ single crystals \cite{Mandal_2021} at 30 K. Total instrumental resolutions were $\sim$300 meV for Al $K_{\alpha}$ (1486.6 eV) and $\sim$10 meV for ARPES using He {\scriptsize I} (21.2 eV) radiations. DFT calculations were performed using projector augmented wave method as implemented in VASP \cite{PhysRevB.54.11169} with plane-wave energy cut-off of 500 eV and $\Gamma$-centered 18 $\times$ 18 $\times$ 10 $k$-mesh, employing the Generalized Gradient Approximation parameterized by Perdew, Burke, and Ernzerhof \cite{PhysRevLett.77.3865}.  Surface band structure  was calculated using Te-terminated slab of ten primitive unit cells stacked along [001] direction with 10 \AA ~of vacuum. SOC was included in all calculations. Detailed descriptions of the experimental and computational methodologies have been provided in the Supplementary Material (SM) \cite{supp}.  
\begin{figure}[t]     \centering\includegraphics[width=.48\textwidth]{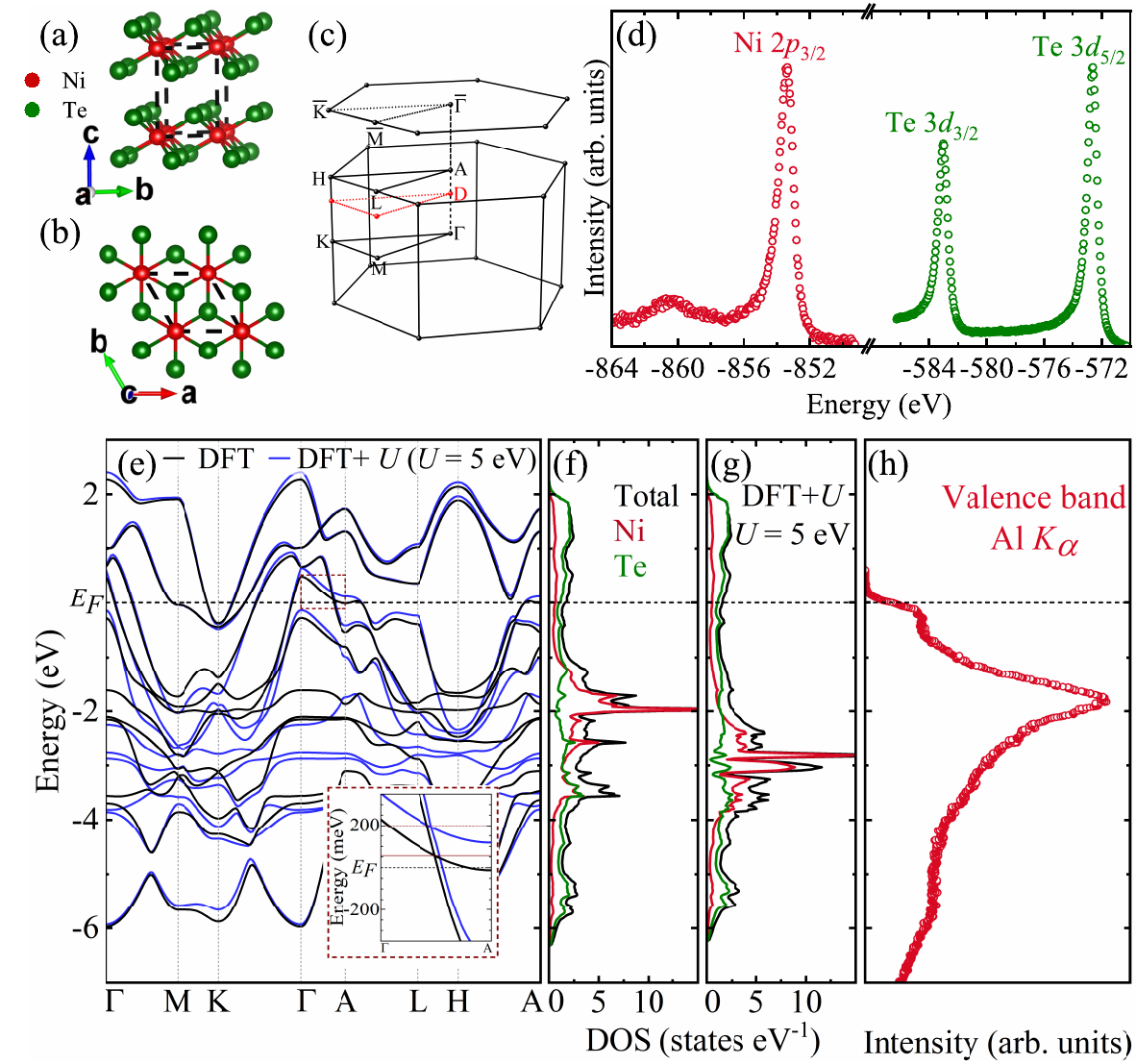}
\vspace{-0.3cm}
\caption{(a) Side and (b) top views of NiTe$_2$ crystal structure. (c) Bulk and surface [001] projected Brillouin zone. Type-II Dirac point appears along $\Gamma$-A direction at D ($k_z \sim0.35c^*$). (d) Core level photoemission spectra of Ni 2$p_{3/2}$ (red) and Te 3$d$ (green). (e) Band structure and (f)-(g) DOS obtained within DFT (black) and DFT+$U$ (blue). Inset shows the enlarged view of Dirac crossing. (h) Valence band photoemission spectra obtained using Al $K_{\alpha}$ radiation.}\label{Fig1}
\end{figure}

NiTe$_{2}$ crystallizes in a trigonal centrosymmetric CdI$_{2}$-prototypical structure (space group $P\overline{3}m1$, no. 164). As illustrated in Fig.~\ref{Fig1}(a), Ni layer is sandwiched between Te layers, with the stacking of adjacent Te layers by weak van der Waals interactions, making [001] plane as a natural cleavage plane leading to the Te terminated surface. The top view in Fig. \ref{Fig1}(b) reveals a hexagonally arranged structure. The corresponding bulk Brillouin zone (BZ) and projected [001] surface BZ, along with high symmetry points are depicted in Fig. \ref{Fig1}(c).

Ni 2$p$ and Te 3$d$ core level photoemission spectra have been shown in Fig. \ref{Fig1}(d). Since the Ni 2$p_{1/2}$ peak overlaps with Te 3$p_{1/2}$ (See Fig. S1 in SM \cite{supp}), we only show the isolated Ni 2$p_{3/2}$ region exhibiting main peak at $\sim$  $-$853.3 eV energy along with a broad satellite appearing at $\sim$ $-$860.3 eV energy. The larger energy separation and smaller intensity of the satellite feature with
respect to the main feature, along with reduced asymmetry of the main feature in comparison to Ni metal is reminiscent of semimetallic character of NiTe$_{2}$ \cite{PhysRevB.61.16370,PhysRevLett.27.479,Nesbitt2000,PhysRevB.22.3644} (see Fig. S1 in SM \cite{supp}). The Te 3$d$ spectral region exhibits two peak structure at $\sim$ $-$572.6 eV and $\sim$ $-$583.0 eV energy corresponding to spin-orbit split 3$d_{5/2}$ and 3$d_{3/2}$ peaks, respectively, with spin-orbit splitting of about 10.4 eV. Sharp core levels, along with the absence of any oxide-related features in the spectral region of Ni and Te suggest a clean sample surface.

   \begin{figure*}[t]
    \centering
    {\includegraphics[width=0.6\textwidth]{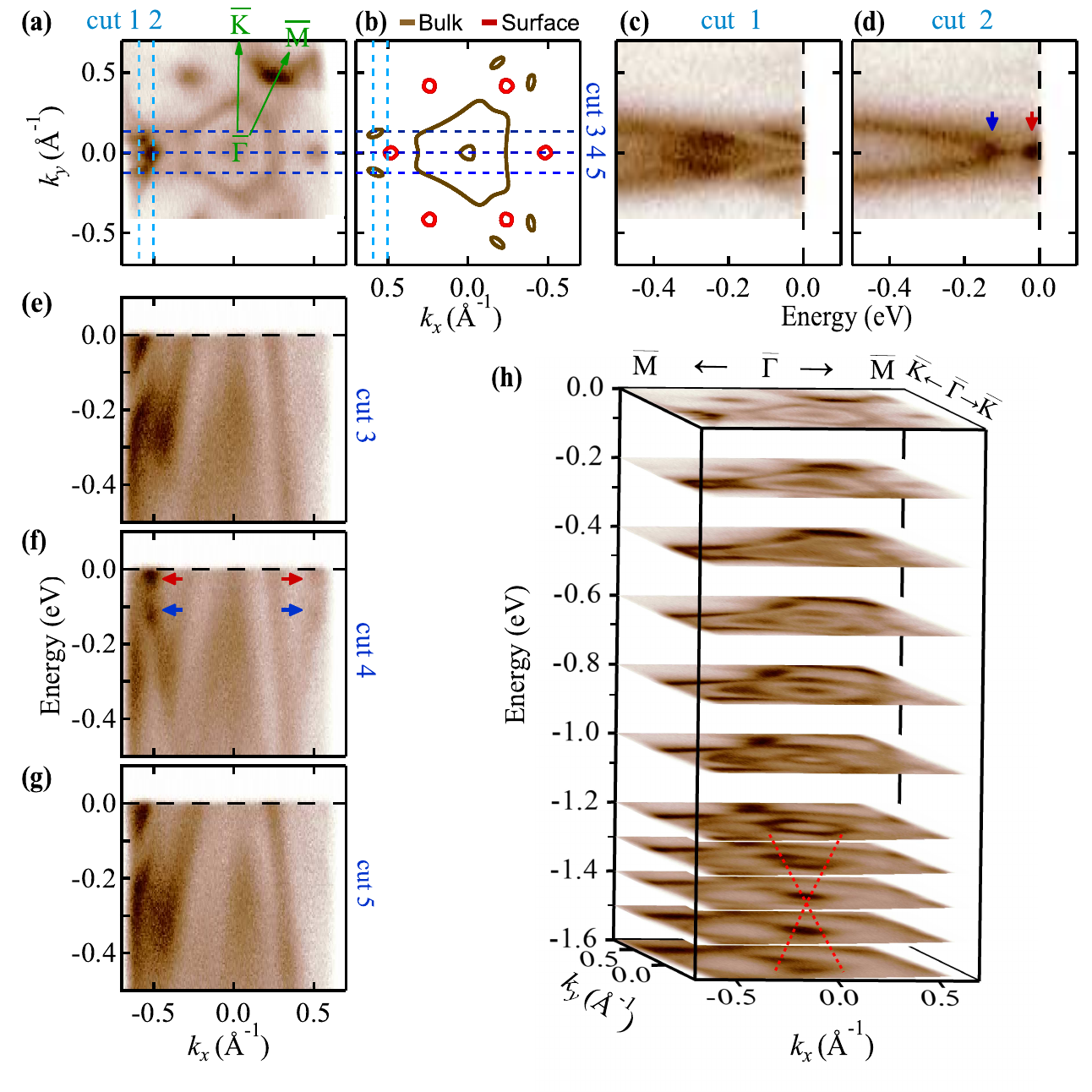}} 
\vspace{-0.1cm}
    \caption{ (a) ARPES FS map and (b) DFT calculated bulk (dark brown) and surface (red) contributions to FS of NiTe$_{2}$. Band dispersion along various momentum cuts for constant (c)-(d) $k_x$ and (e)-(g) $k_y$. Red and blue arrows mark the topological surface states. (h) Three-dimensional plot of constant energy maps from 0 eV to $-$1.6 eV energy from the ARPES spectra. Red dotted lines are indicative of the Dirac crossing formed by SS2.}\label{Fig2}
    \end{figure*}

 To understand the electronic structure of NiTe$_2$, we show the DFT calculated bulk band dispersion by black lines in Fig.~\ref{Fig1}(e) and the corresponding DOS in Fig. \ref{Fig1}(f). The dominant Ni 3$d$ contribution appears at around $-$2 eV energy, while the states in the vicinity of $E_F$ have strongly hybridized character. As evident, four bands cross $E_F$, giving rise to two electron pockets centered around the K point and two hole pockets centered around $\Gamma$ point.  The corresponding atom projected band structure and FSs have been shown in Fig. S2 and S3, respectively, in the SM \cite{supp}. Interestingly, the bands forming hole pockets intersect along the ${\Gamma}$-A direction and give rise to type-II Dirac crossing. The titled Dirac cone appears at $\sim$55 meV energy and $k_{z}$ = $\sim$0.35 $c^*$ ($c^*=2\pi/c$). The energy position of the Dirac crossing crucially depends on the structural parameters, and it is possible to tune the type-II Dirac point to $E_F$, making it a suitable platform for various applications \cite{PhysRevB.103.125134,PhysRevB.110.L201401}. The total DOS at $E_F$ is found to be 1.67 states/eV, which is about 15\% smaller than the value obtained from specific heat measurements suggesting weak correlation in NiTe$_2$ \cite{Mandal_2021}. To further understand the role of electron correlation we show the band dispersion calculated within DFT+$U$ method using blue lines in Fig. \ref{Fig1}(e), and the corresponding DOS is presented in Fig. \ref{Fig1}(g), where the overall band structure remains similar around $E_F$ with a shift of the Dirac point towards the higher energy. A moderate value to $U$ = 5 eV shifts the Dirac point to 200 meV (see inset) while the Ni dominated states move away from $E_F$ and appear around $-$3 eV energy (See Fig. S4 in SM for DOS obtained for different $U$ as well as results obtained using meta-GGA (SCAN) and hybrid (HSE06) functionals \cite{supp}).

For a direct comparison with theoretical results, the bulk-sensitive valence band spectra using Al $K_{\alpha}$ radiation has been shown in Fig. \ref{Fig1}(h). The spectra reveal a distinct feature at approximately $-$2 eV, accompanied by shoulder structures ranging from $-$2.5 eV to $-$4 eV, and a broad feature around $-$5.5 eV. The presence of a small and constant spectral intensity near $E_F$, along with the Fermi cutoff indicates a semimetallic nature. Notably, the spectra primarily represent the Ni 3$d$ states, as the photoionization cross-section for Ni 3$d$ states is significantly larger ($\sim$2.5 times) than that for Te 5$p$ states for 1486.6 eV photon energy \cite{YEH19851}. As evident, various features in experimental valence band spectra, including Te dominant region appearing at around $-$5.5 eV energy are very well reproduced in the DOS obtained within DFT. Contrasting to experimental observation, the shift of the dominant Ni 3$d$ states towards lower energy in DOS obtained within DFT+$U$ suggests that the electron correlation effects are unimportant in the description of bulk electronic structure of NiTe$_{2}$, as also reported earlier \cite{Mukherjee2020}.  

   \begin{figure*}[t]
    \centering
    {\includegraphics[width=.75\textwidth]{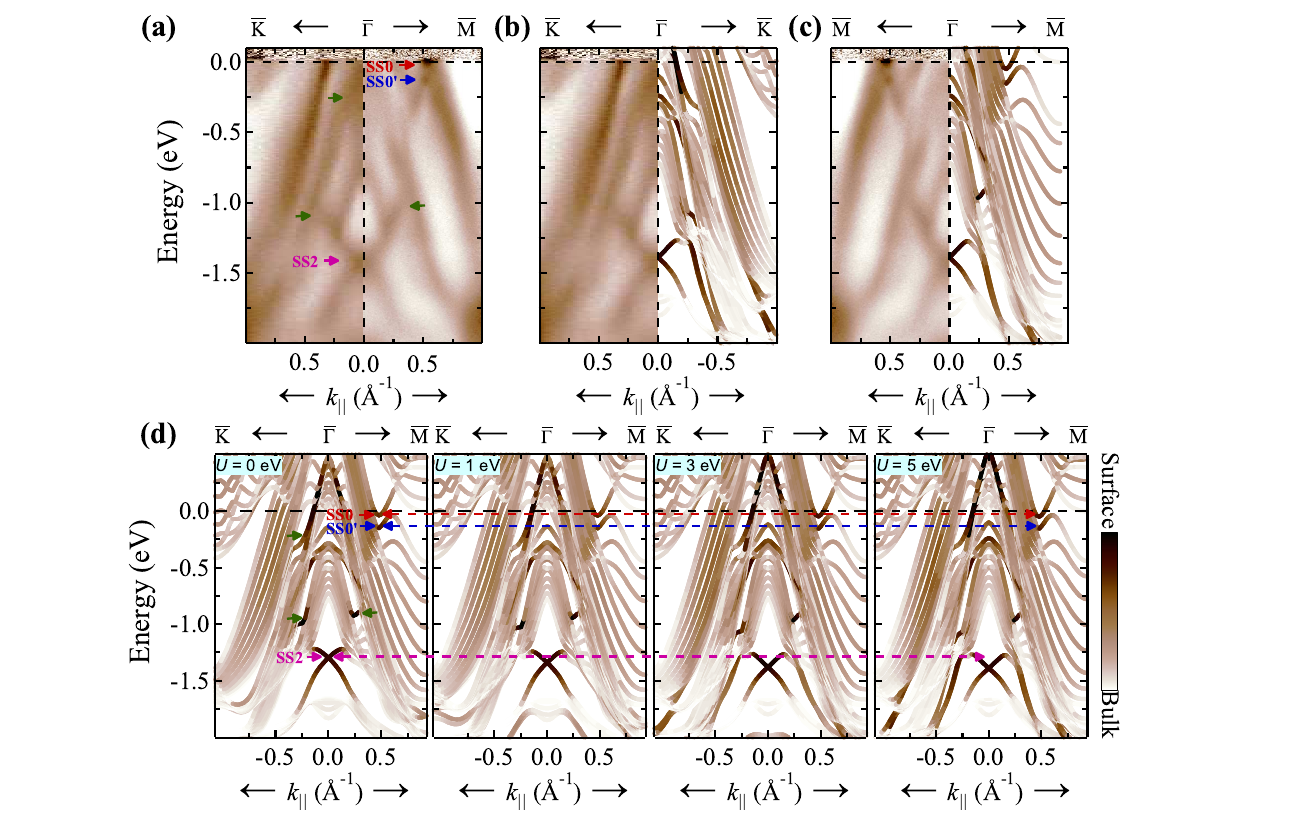}}
\vspace{-0.3cm}
    \caption{(a) ARPES spectra of NiTe$_2$ along $\overline{\textrm{K}}$-$\overline{\Gamma}$-$\overline{\textrm{M}}$ direction. Comparison of ARPES (left panel) and DFT+$U$ ($U$ = 5 eV) (right panel) results along (b) $\overline{\Gamma}$-$\overline{\textrm{K}}$ and (c) $\overline{\Gamma}$-$\overline{\textrm{M}}$ directions. (d) Surface band structure calculated within DFT+$U$ method.} \label{Fig3}
    \end{figure*} 

To further investigate the electronic band dispersion and the surface states, we show the results of ARPES spectra obtained using He {\scriptsize I} radiation in Fig. \ref{Fig2}. As shown in Fig. \ref{Fig2}(a), the constant energy map at 0 eV clearly reveals a three-fold symmetric FS. A small circular and an irregular hexagon-shaped FS with 3-fold symmetry, centered at $\overline{\Gamma}$, is clearly evident. Additionally, we observe an arc-shaped feature and a circular feature appearing in the alternate $\overline{\Gamma}$-$\overline{\textrm{M}}$ direction. A closer look reveals that the arc-shaped feature consists of three small pockets. The FS of NiTe$_2$ is highly anisotropic, exhibiting strong dispersion along the $k_z$ direction despite the two-dimensional (2D) nature of the crystal structure (see Fig. S3 of SM for DFT calculated 3D FSs and 2D cuts \cite{supp})). The 21.2 eV photon energy used in the present case probes the $k_z$ = 0.34 $c^*$ plane of the BZ for NiTe$_2$ \cite{PhysRevB.100.195134}. Consequently, we present the DFT calculated bulk FS cut for the corresponding $k_z$ value in Fig. \ref{Fig2}(b) with dark brown lines showcasing the $\overline{\Gamma}$ centered two hole pockets along with pair of electron pockets in the alternate $\overline{\Gamma}$-$\overline{\textrm{M}}$ direction. The bulk FS well replicate the ARPES features with the exception of six-fold symmetric circular features along $\overline{\Gamma}$-$\overline{\textrm{M}}$ direction. As reported earlier, NiTe$ _2 $ exhibits various surface states; among them, two of the surface states have been extensively discussed in the literature, (\textit{i}) parabolic-shaped SOC split bands SS0 and SS0$^{\prime}$ appearing along the $\overline{\Gamma}$-$\overline{\textrm{M}}$ direction (red and blue arrows in Fig. \ref{Fig2}(d) and \ref{Fig2}(f)) intersecting $E_F$ to form electron pockets, and (\textit{ii}) a Dirac-like conical crossing at $\overline{\Gamma}$ point formed by surface state SS2 within the parity inverted bulk band gap much below $E_F$ \cite{PhysRevB.100.195134,PhysRevB.104.155133,Mukherjee2020}. The calculated surface FS using slab configuration within DFT exhibits six fold symmetric circular electron pockets along $\overline{\Gamma}$-$\overline{\textrm{M}}$ direction, which have been shown by red lines in Fig. \ref{Fig2}(b). The overall experimental FS aligns excellently with the DFT results shown in Fig. \ref{Fig2}(b) indicating a three-fold symmetry of bulk and six-fold symmetry of surface state. The momentum broadening in the experimental spectra results in a visual appearance of arc-shaped FS due to the presence of  three nearby electron pockets in the alternate $\overline{\Gamma}$-$\overline{\textrm{M}}$ direction.  

In Fig. \ref{Fig2}(c-g), we show the band dispersions along various momentum cuts (dashed lines in the FS map) for a clear distinction between the surface states SS0 and SS0$^{\prime}$ with the bulk states. The constant $k_x$ cuts 1 and 2 (cyan dashed lines) are taken for $k_x $ = $-$0.6 \AA$^{-1}$ and $-$0.5 \AA$^{-1}$ with corresponding band dispersions shown in (c) and (d), respectively. The band dispersion along cut 1 exhibits two bulk bands crossing $E_F$ at  $k_y$ = $\sim$ $\pm$0.1 \AA$^{-1}$ and additional bulk bands dispersing at lower energies. While the dispersion along cut 2 exhibits two topological surface states (SS0 \& SS0$^{\prime}$) within $\sim$0.2 eV below $E_F$ around $k_y$ = 0 \AA$^{-1}$, as also observed from DFT (discussed later). Very similar band dispersion along cuts 3 \& 5 at constant $k_y$ = $\pm$0.1 \AA$^{-1}$ indicate bulk band symmetry about $k_y$ = 0 \AA$^{-1}$. Band dispersion along cut 4 exhibit SS0 and SS0$^\prime$ dispersing upto $\sim$0.2 eV below $E_F$ at around $k_x$ = $\pm$0.5 \AA$^{-1}$. Due to the matrix element effect the asymmetric distribution of spectral intensity of the surface states can be clearly seen in two directions away from $\overline{\Gamma}$ point as also observed earlier \cite{Mukherjee2020}.

Further, we show the constant energy cuts from the ARPES spectra in the wide energy range as a three-dimensional plot in Fig. \ref{Fig2}(h). It is important to note the appearance of the SS2 as conical shaped band (Indicated by red dotted line) in the energy range $-$1.2 eV to $-$1.6 eV with Dirac-like crossing at $\sim$ $-$1.4 eV (shown with 0.1 eV energy step for constant energy cuts). The front view of the constant energy maps have been shown in Fig. S5, in the SM \cite{supp}.

So far, the FS and states close to $E_F$ have been discussed, and we now focus on the Dirac-like crossing of the topological surface state SS2 appearing at  $-$1.42 eV in the ARPES spectra. The position of this crossing was found to be $-$1.30 eV in our DFT calculated surface band structure. This discrepancy has also been observed in various reports and have been overlooked where the experimentally observed crossing always appears at lower energy than the DFT obtained results \cite{PhysRevB.100.195134,PhysRevB.104.155133,PhysRevB.110.L201401}. A recent report highlights strain-induced modification of surface states where both ARPES and DFT results reveal a lower energy shift of SS2 as well as SS0 \& SS0$^{\prime}$, suggesting a critical role of lattice parameters in altering the topological surface states \cite{PhysRevB.110.L201401}.
The consistent discrepancy of the lowered energy position of SS2 in  ARPES experiments compared to various theoretical results, despite having good agreement for SS0 \& SS0$^{\prime}$ \cite{PhysRevB.104.155133,Mukherjee2020}, indicates a missing ingredient in theoretical approaches and warrants further investigation. Chalcogen vacancies can have significant impact on the electronic properties of TMDCs  \cite{PhysRevB.92.235408,Zhussupbekov2021}. For example, selenium vacancy on the surface significantly modifies the topological surface states in Bi$_2$Se$_3$ \cite{https://doi.org/10.1002/pssr.201206415}.
Such effects can be ruled out in NiTe$_2$ as our theoretical results incorporating surface chalcogen vacancies exhibit a higher energy shift of SS2 as opposed to the experimental observation (detailed discussion in SM \cite{supp}).

The surface state SS2 is formed within the parity inverted band gap formed by the bands having hybridized Ni 3$d$ and Te 5$p$ states \cite{Bahramy2018,Mukherjee2020}. 
In the present case, surface state having substantial 3$d$ character may get affected due to electronic correlation ( See Fig. S7 in SM for hybridization, wavefunction plot and effect of $U$ on SS2 \cite{supp}).
The ARPES study and subsequent theoretical work demonstrate the importance of electronic correlations in MoTe$_{2}$ leading to  renormalization of  electronic bands \cite{PhysRevB.99.035123,PhysRevLett.121.136401}. However, our  bulk-sensitive valence band spectra as well as ARPES FS, show good agreement with DFT results for NiTe$_2$. Electron correlation effects are significantly enhanced  at the surface than the bulk, where the effective correlation strength is expected to be larger as a result of significantly smaller screening and coordination number \cite{K.Maiti_2001,PhysRevB.71.161102,PhysRevLett.90.096401,N.Kamakura_2004,Ali_2023}. Our constrained random phase approximation (cRPA) 
\cite{PhysRevB.80.155134} calculations to obtain the Hubbard $U$ (details in SM \cite{supp}) leads to screened $U$ of 2.17 eV for the bulk with moderately increased value of 2.61 eV for the monolayer of NiTe$_2$ \cite{PhysRevB.105.115115}, indicating an enhancement of electron correlation at the surface. Thus, we calculate the surface band structure of NiTe$_2$ within DFT+$U$ method, with $U$ applied to the surface Ni atoms only and compare the results with ARPES.

Fig. \ref{Fig3}(a) shows the  ARPES spectra along $\overline{\textrm{K}}$-$\overline{\Gamma}$-$\overline{\textrm{M}}$ direction (green lines shown in Fig. \ref{Fig2}(a)). The red \& blue and pink arrows highlight the surface states SS0 \& SS0$^{\prime}$ and SS2, respectively (see Fig. S8 in SM for ARPES spectra along other momentum directions \cite{supp}). As identified experimentally in the literature \cite{PhysRevB.104.155133}, location of few other prominent surface states are also marked using green arrows. The DFT+$U$ band dispersion using slab configuration for varying values of $U$ have been shown in various panels of Fig. \ref{Fig3} (d). Here, the dark brown colors (in the intensity map) represent the surface states and have been similarly marked in the first panel corresponding to DFT results ($U$ = 0). The position of the Dirac-like crossing at $-$1.30 eV is marked by the dashed pink horizontal line while the SS0 \& SS0$^{\prime}$ have been marked by red \& blue horizontal lines, respectively. As evident in Fig. \ref{Fig3}(d), inclusion of electron correlation to surface Ni atoms leads to a monotonous decrease in the energy position of SS2 while all other surface states remain mostly unaffected. Notably, for $U$ = 5 eV, SS2 shifts by $\sim$100 meV, while SS0 \& SS0$^{\prime}$ remains unaffected.  
 Thus, for a one to one correspondence, we show the  results for $U$ = 5 eV alongside with ARPES spectra along $\overline{\Gamma}$-$\overline{\textrm{K}}$ and $\overline{\Gamma}$-$\overline{\textrm{M}}$ in Fig. \ref{Fig3}(b) and \ref{Fig3}(c), respectively. The striking similarity between the experimental spectra and theoretical results with respect to the energy dispersion and positions of surface state SS2 forming Dirac-like conical crossing as well as SS0 \& SS0$^{\prime}$ can be clearly seen. Our results demonstrate that the inclusion of on-site Coulomb interaction is essential for a precise description of the surface electronic structure of NiTe$_2$.

In summary, we highlighted the role of electron correlation in the surface electronic structure of NiTe$_{2}$ using photoemission spectroscopy and band structure calculation within DFT and  DFT+$U$ theoretical approach. The experimental valence band and Fermi surface can be well captured within the DFT framework. The close proximity of surface state along $\overline{\Gamma}$-$\overline{\textrm{M}}$ direction to $E_F$ distinguishes it from other surface states and suggests a finite contribution of topological surface carriers to the non-trivial transport. Topological surface state formed by the Dirac-like conical crossing at energy $-$1.42 eV in the ARPES spectra show a discrepancy with DFT results. The electron correlation effects are enhanced while going from bulk to surface and hybridization of the Ni $3d$ and Te $5p$ states results in electron correlation induced shifting of the conical surface state. Our study demonstrates that properly treating the electronic correlation is crucial for accurately determining the band topology and topological surface state of NiTe$_{2}$. 

 N.B. and D.S. acknowledge the CSIR India, for financial support through Awards No. 09/1020(0177)/2019-EMR-I and 09/1020(0198)/2020-EMR-I, respectively. R.P.S. acknowledges the SERB India, for Core Research Grant No. CRG/2019/001028. We gratefully acknowledge the use of the HPC facility and CIF at IISER Bhopal.

\end{document}


\title{Supplementary Material for ``Strongly correlated topological surface states in type-II Dirac semimetal NiTe$_{2}$"}
\author{Neeraj Bhatt}
\affiliation{Department of Physics, Indian Institute of Science Education and Research Bhopal, Bhopal Bypass Road, Bhauri, Bhopal 462066, India}%

\author{Asif Ali}
\affiliation{Department of Physics, Indian Institute of Science Education and Research Bhopal, Bhopal Bypass Road, Bhauri, Bhopal 462066, India}%

\author{Deepali Sharma}
\affiliation{Department of Physics, Indian Institute of Science Education and Research Bhopal, Bhopal Bypass Road, Bhauri, Bhopal 462066, India}%

\author{Sakshi Bansal}
\affiliation{Department of Physics, Indian Institute of Science Education and Research Bhopal, Bhopal Bypass Road, Bhauri, Bhopal 462066, India}%

\author{Manasi Mandal}
\affiliation{Department of Physics, Indian Institute of Science Education and Research Bhopal, Bhopal Bypass Road, Bhauri, Bhopal 462066, India}%

\author{Ravi Prakash Singh}
\affiliation{Department of Physics, Indian Institute of Science Education and Research Bhopal, Bhopal Bypass Road, Bhauri, Bhopal 462066, India}%

\author{Ravi Shankar Singh}
\email{rssingh@iiserb.ac.in}
\affiliation{Department of Physics, Indian Institute of Science Education and Research Bhopal, Bhopal Bypass Road, Bhauri, Bhopal 462066, India}%

\date{\today}

\maketitle

\section{Experimental and computational details }
Well characterized single crystals of NiTe$_{2}$ \cite{Mandal_2021} were cleaved \textit{in-situ} (base pressure $<$ 4 $\times$ 10$^{-11}$ mbar) at 30 K for all the photoemission spectroscopic measurements using Scienta R4000 electron analyzer and monochromatic photon sources. Total instrumental resolutions were set to $\sim$300 meV for Al $K_{\alpha}$ (1486.6 eV) and $\sim$10 meV for angle resolved photoemission spectroscopy (ARPES) using He {\scriptsize I} (21.2 eV) radiations (energy). The Fermi level (E$_F$) and total energy resolutions were determined by measuring the Fermi edge of a clean polycrystalline Ag at the same temperature. The samples were mounted on a 5-axis sample manipulator, and the analyzer slit was parallel to the $M$-$\Gamma$-$M$ direction while the sample was rotated along $K$-$\Gamma$-$K$ direction (perpendicular to the slit) in the steps of 0.5$^\circ$ ($\sim$ 0.018 \AA$^{-1}$) for ARPES data.

Density functional theory (DFT) calculations were performed using projector augmented wave method as implemented in VASP \cite{PhysRevB.54.11169} with plane-wave energy cut-off of 500 eV and $\Gamma$-centered 18 $\times$ 18 $\times$ 10 $k$-mesh, employing the Generalized Gradient Approximation (GGA) parameterized by Perdew, Burke, and Ernzerhof (PBE) \cite{PhysRevLett.77.3865}. Full optimization of the lattice structure (atomic forces below 0.5 meV\AA$^{-1}$) led to the relaxed lattice parameters $a = b =$ 3.897 \AA, $c =$ 5.284 \AA; which closely match with the experimental values \cite{Xu2018}. DFT calculations using meta-GGA with strongly constrained and appropriately normed (SCAN) functional \cite{PhysRevLett.115.036402} and screened hybrid functional of Heyd, Scuseria, and Ernzerhof (HSE06) \cite{10.1063/1.1564060} were also performed using a $\Gamma$-centered 12 $\times$ 12 $\times$ 6 $k$-mesh. Surface band structure calculations within DFT (PBE) were performed using a Te-terminated slab of ten primitive unit cells stacked along [001] direction with 10 \AA ~of vacuum. Spin-orbit coupling was included in all calculations. We have also performed constrained random phase approximation (cRPA) \cite{PhysRevB.80.155134} calculations to estimate the Hubbard $U$ parameter for the bulk as well as for the monolayer of NiTe$_2$. The polarization function was evaluated by constructing Ni 3$d$ and Te 5$p$ maximally localized Wannier functions \cite{PhysRevB.80.155134} using WANNIER90 program \cite{MOSTOFI2008685} and only Ni 3$d$ orbitals were considered in the correlated subspace. The screened $U$ = 2.17 eV and 2.61 eV were obtained for bulk and monolayer, for the static limit ($\omega$ = 0) of the cRPA interactions \cite{PhysRevB.105.115115}. Important to note here that the Hubbard $U$ parameter can also be computed using linear response theory (LRT) \cite{LRT} or density functional perturbation theory (DFPT) \cite{HP-DFPT} where the convergence must be achieved with respect to real space (supercell size) or momentum space ($q$-mesh) respectively. In all the three methodologies (cRPA, LRT and DFPT), obtaining $U$ for different Ni atoms (surface and bulk) in the 10 unit cell thick slab configiration requires forbiddingly expensive computation and is beyond the scope of our computational resources.

  \section{X-ray photoemission survey scan spectra}  
   
\begin{figure*}[t]
 	\centering
 	{\includegraphics[width=1.00\textwidth]{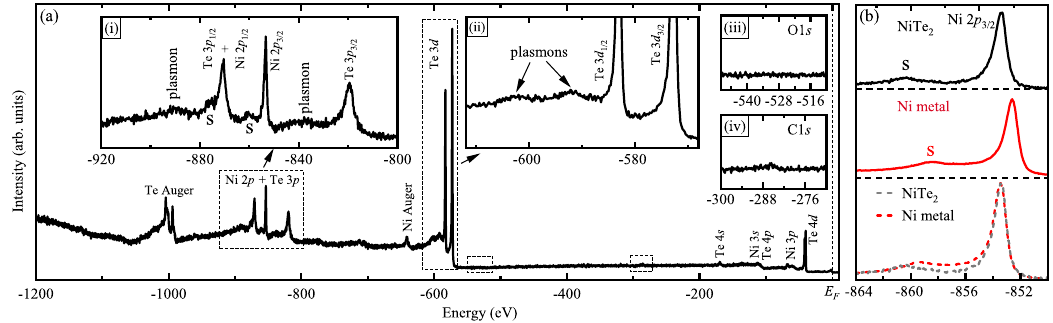}}
 	\caption{(a) XPS survey spectra of NiTe$_{2}$. Insets (i)-(iv) show the enlarge view of energy regions corresponding to Ni 2$p$ + Te 3$p$, Te 3$d$, O 1$s$ and C 1$s$ core level spectra, respectively. Plasmon loss features corresponding to Te core level have also been indicated. (b) Comparison of Ni 2$p_{3/2}$ main and satellite peaks of NiTe$_{2}$ and pure Nickle metal. }\label{FigS1}
 \end{figure*}

 Survey scan of NiTe$_{2}$ collected using Al $K_\alpha$ radiation at 30~K exhibits various features as marked in Fig. \ref{FigS1}(a). The insets show an enlarged view of the spectral region corresponding to (i) Ni 2$p$ + Te 3$p$, (ii) Te 3$d$, (iii) O 1$s$ and (iv) C 1$s$ core levels. Absence/negligibly small intensity corresponding to oxygen and carbon features in the survey scan spectra confirms an extremely clean sample surface. Multiple samples were cleaved \textit{in-situ} to ensure the cleanliness and reproducibility of the X-ray photoemission spectra (XPS) and angle resolved photoemission spectra (ARPES).

The Ni 2$p$ + Te 3$p$ spectral region exhibits multiple features, as shown in the inset (i). The Te 3$p_{3/2}$ and  Ni 2$p_{3/2}$ peaks appear at  $-$819.6 eV and $-$853.4 eV, respectively,  while their corresponding spin-orbit split Te 3$p_{1/2}$ and Ni 2$p_{1/2}$ overlap and appears at $\sim$ $-$870.6 eV, suggesting spin-orbit splitting of about  51.0 eV and 17.2 eV for Te 3$p$  and Ni 2$p$, respectively. The famous satellite features (marked as 'S') corresponding to Ni 2$p$ are also distinctly visible in the spectra. Additionally, broad features appearing at $\sim$ $-$837.6 eV and $\sim$ $-$888.6 eV correspond to plasmon satellites related to Te 3$p_{3/2}$ and Te 3$p_{1/2}$, respectively. The plasmon satellites appearing at $\sim$18 eV lower energy than the respective core levels are also seen in the Te 3$d$ spectral region as marked by arrows in the inset (ii).  The satellite feature S in the Ni 2$p$ core level spectra have been observed in various nickel-based systems and provide valuable information related to interactions parameters and electronic structure \cite{PhysRevB.61.16370,PhysRevLett.27.479}. Therefore, we compare the Ni 2$p_{3/2}$ spectra  of NiTe$_2$ (black line) and pure Ni metal (red line) in Fig. \ref{FigS1}(b). The main feature exhibits an asymmetric lineshape and appears at $\sim$ $-$853.4 eV and $\sim$ $-$852.6 eV in case of NiTe$_2$ and Ni, respectively. Interestingly, the energy separation of the satellite feature S with respect to the main feature is $\sim$7 eV in the case of NiTe$_{2}$, while it is 6 eV for pure Ni metal. It is already established that the presence and nature of the ligand significantly influences the position and intensity of the satellite feature \cite{Nesbitt2000}. For a direct comparison, we shift the Ni spectra (0.8 eV) to match the energy position of the main peak as shown in bottom panel. The lower asymmetry of the main peak and  larger separation with smaller intensity of the satellite feature indicates the semimetallic  nature of NiTe$_{2}$ \cite{PhysRevB.22.3644,Nesbitt2000}.

  \begin{figure}[b]
    \centering
    {\includegraphics[width=0.47\textwidth]{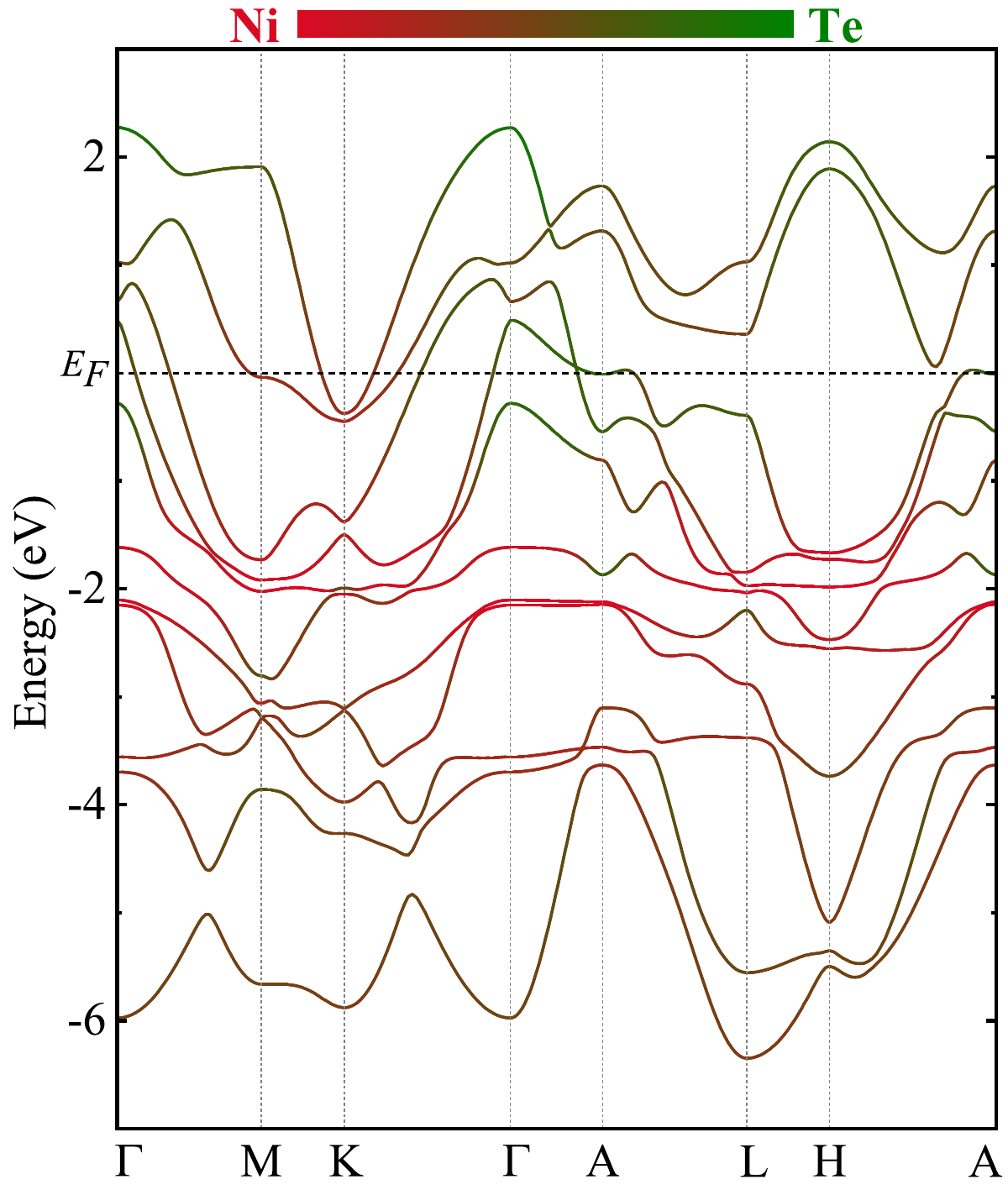}}
    \caption{ DFT calculated atom projected band structure.}  \label{FigS2}
    \end{figure}

 \begin{figure*}[t]
    \centering
{\includegraphics[width=0.93\textwidth]{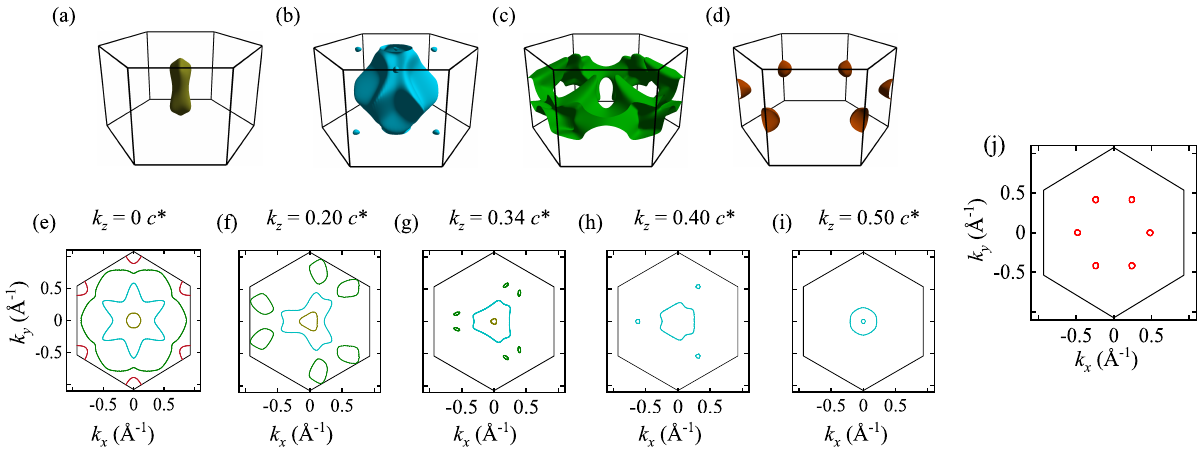}}
    \caption{ DFT Calculated (a)-(d) individual bulk 3D Fermi surface (FS) of four bands crossing the Fermi level and corresponding (e)-(i) in plane  ($k_{x}$-$k_{y}$) FS cut for different $k_{z}$ values ranging from the center (${\Gamma}$) to the boundary (A) of BZ. (j) DFT calculated surface FS derived from the surface state SS0.}\label{FigS3}
    \end{figure*}

\section{DFT Calculated band structure and Fermi surfaces}
We show the atom projected band structure calculated using DFT in Fig. \ref{FigS2} with Ni and Te atomic contributions highlighted by red and green colors, respectively. The states near the E$_F$ exhibit strong hybridization between Te 5$p$ and Ni 3$d$ orbitals, while Ni-dominated states appear around $-$2 eV.  As discussed in the main text, the intense peak centered at $-$2 eV in the valence band spectra primarily corresponds to Ni 3$d$ states. 

As seen in Fig. \ref{FigS2} (also discussed in the main text), four bands cross the $E_F$, forming  highly anisotropic FSs, which have been shown in Fig. \ref{FigS3}. There are spindle-shaped (dark yellow) and  hexagonal barrel-shaped (cyan) hole-like FSs, as shown in the Fig. \ref{FigS3}(a) and Fig. \ref{FigS3}(b) respectively. 
 A gear-shaped (green) and a bowl-shaped (dark brown) electron-like FSs have been shown in Fig. \ref{FigS3}(c) and Fig. \ref{FigS3}(d) respectively. Despite having a two-dimensional (2D) nature of the crystal structure, the FS is evidently anisotropic along the $k_z$ direction, thus, we show the 2D FS cuts for different $k_z$ values from the Brillouin zone (BZ) center (${\Gamma}$) to the BZ boundary (A) in Fig. \ref{FigS3}(e), (f), (g), (h) and (i).  The symmetry of the FS is six-fold at ${\Gamma}$ and A, while it shows a three-fold symmetry for other $k_z$ values. At the BZ center ($k_z$ = 0), 2D FS cut exhibits a small circular shaped along with a star-shaped hole pocket centered at ${\Gamma}$. The hexapetalus flower-shaped electron pocket is also centered at ${\Gamma}$, while the bowl-shaped electron pocket appears at each K point (Fig. \ref{FigS3}(e)). With increasing $k_z$ the star and  hexapetalus contours transform to a three-fold symmetric irregular hexagon and a trifolium-shaped contours while the central circle obtains a triangular form and electron pockets at K point completely disappear. Finally, at BZ boundary the FS cut exhibits two six-fold symmetric closed curve shaped feature centered at ${\Gamma}$ (Fig. \ref{FigS3} (i)).
 Since NiTe$_{2}$ exhibits various topological surface states where the topological surface state, SS0, crosses the E$_F$ \cite{PhysRevB.100.195134,PhysRevB.104.155133,Mukherjee2020}. Thus, to capture the surface FS, we compute the surface electronic structure using the slab configuration within DFT and extract the contribution of topological surface state SS0, which has been shown in Fig. \ref{FigS3} (j) exhibiting a six-fold symmetric circular electron pocket along the $\overline{\Gamma}$-$\overline{\textrm{M}}$ direction. Bulk FS for $k_z$ = 0.34$c^*$ (Fig. \ref{FigS3}(g)) along with surface FS (Fig. \ref{FigS3}(j)) have been overlapped to compare with ARPES FS in the main text (Fig. 2).

      \begin{figure}[b]
	\centering
	{\includegraphics[width=0.305\textwidth]{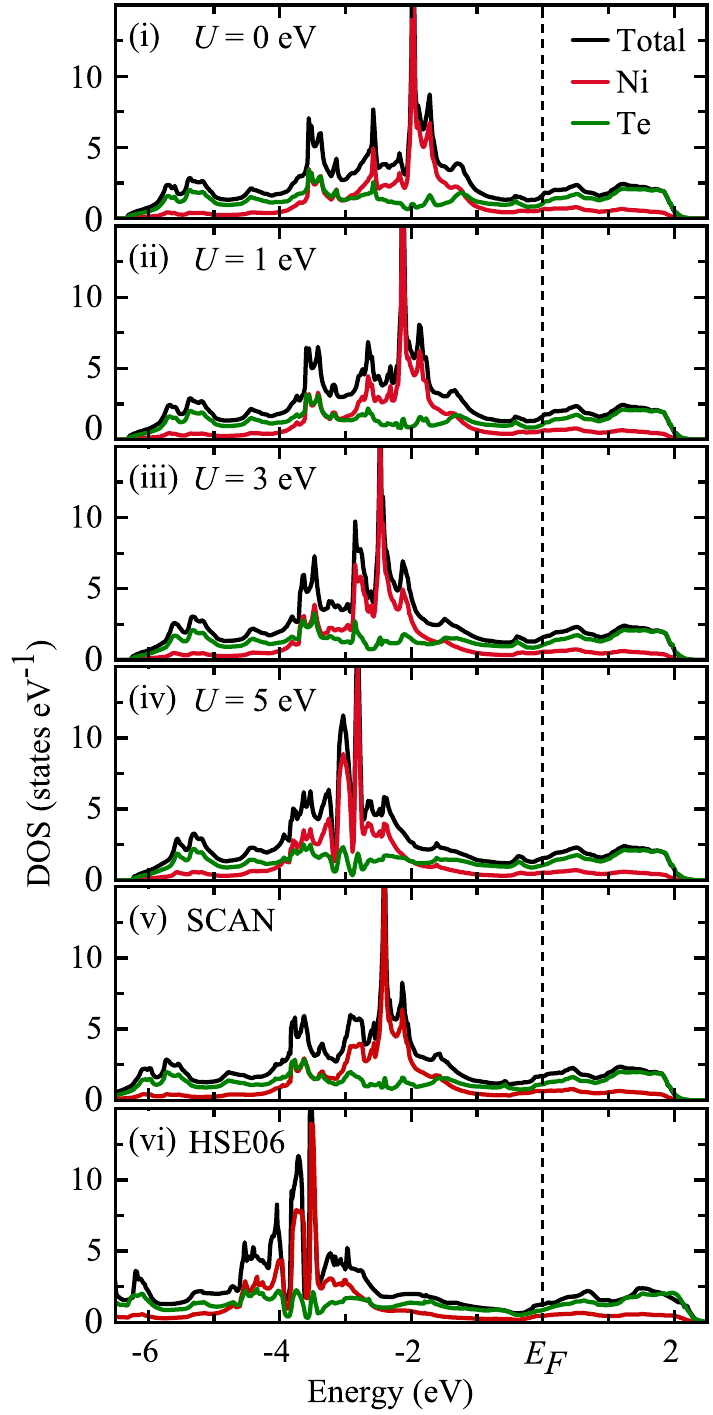}}
	\caption{ Total and partial DOS Calculated using DFT(PBE)+$U$ method with $U$ values of (i) 0 eV, (ii) 1 eV, (iii) 3 eV and (iv) 5 eV. Total and partial DOS calculated using (v) SCAN and (vi) HSE06 functional. }  \label{FigS4}
\end{figure}

\section{ Calculated density of states}

To investigate the role of electron correlation in the bulk electronic structure, we show the results of DFT+$U$ calculations in Fig. \ref{FigS4}. Total and partial density of states (DOS) calculated for $U$ values of 0 eV, 1 eV, 3 eV and 5 eV (for Ni 3$d$) have been shown in Fig. \ref{FigS4}(i), (ii), (iii) and (iv) respectively. As Te 5$p$ states contribute predominantly around $-$5.5 eV and in the close vicinity of $E_F$, the effect of electron correlation remain insignificant in this energy region. The Ni-dominated states, centered at around $-$2 eV for $U$ = 0, monotonously shifts away from the $E_F$ with increasing value of $U$. The total and partial DOS obtained using SCAN and HSE06 functionals have also been shown in Fig. \ref{FigS4}(v) and (vi) respectively. The DOS obtained using SCAN is very similar to that obtained using DFT(PBE)+$U$ (for $U$ = 3 eV), where the Ni 3$d$ PDOS (red line) is centred around $-$2.5 eV in contrast to the bulk-sensitive experimental valence band spectra where Ni dominant feature appears around $-$2 eV energy. DOS obtained using HSE06 leads to a much larger shift of Ni 3$d$ states. These results suggests that DFT using PBE functional accurately describes the bulk electronic structure and confirm that the electron correlation effects are insignificant in the bulk.

      \begin{figure*}[t]
	\centering
	{\includegraphics[width=0.85\textwidth]{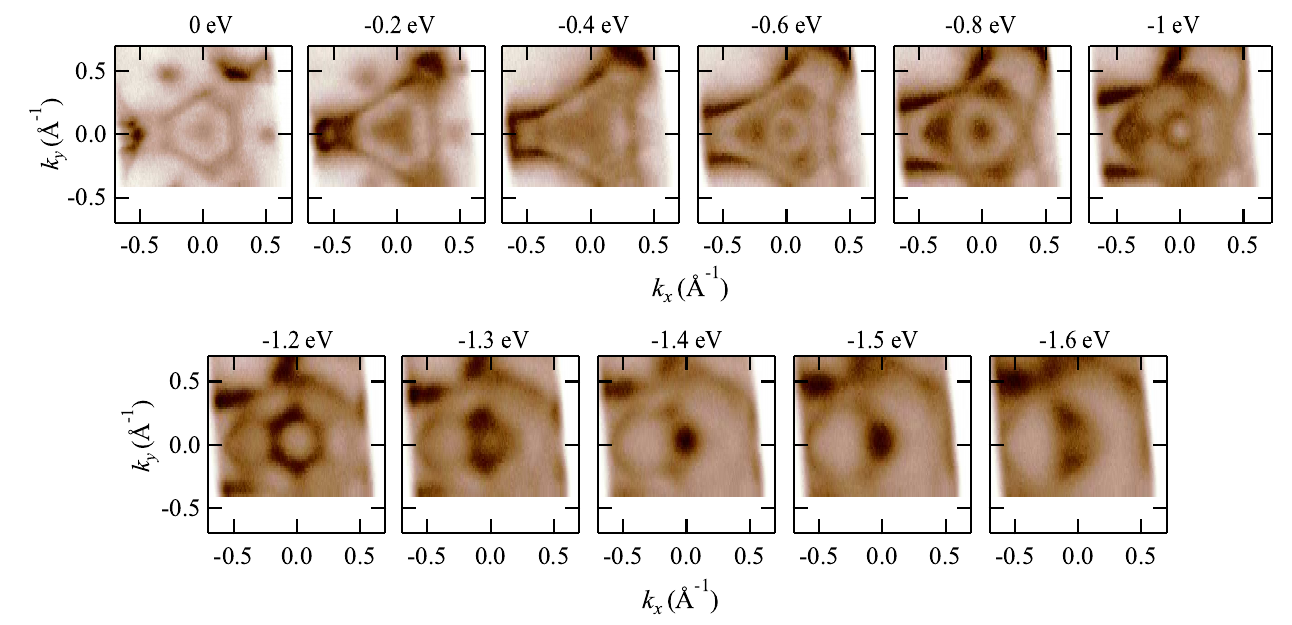}} 
	\caption{  Constant energy maps of the ARPES spectra for different energy.} \label{FigS5}
\end{figure*}

\section{ Constant energy maps of ARPES spectra }
Fig. S5 shows the   constant energy maps from the ARPES spectra, to visualize the evolution of various bulk and surface band features. These maps are same as 3D constant energy plot shown in the Fig. 2(h) of the main text. As evident, multiple bands contribute and disperse in a large energy range, the bulk bands giving rise to electron pockets disappear at 0.2 eV below $E_F$ while $\overline{\Gamma}$ centered hole pockets grow with lowering energy. The surface states SS0 \& SS0$^{\prime}$ disappear beyond 0.2 eV below $E_F$. The topological surface state SS2 dispersing between $-$1.2 eV to $-$1.6 eV, giving rise to Dirac crossing, remain well separated from the bulk bands.

\begin{figure}[b]
	\centering
	{\includegraphics[width=0.48\textwidth]{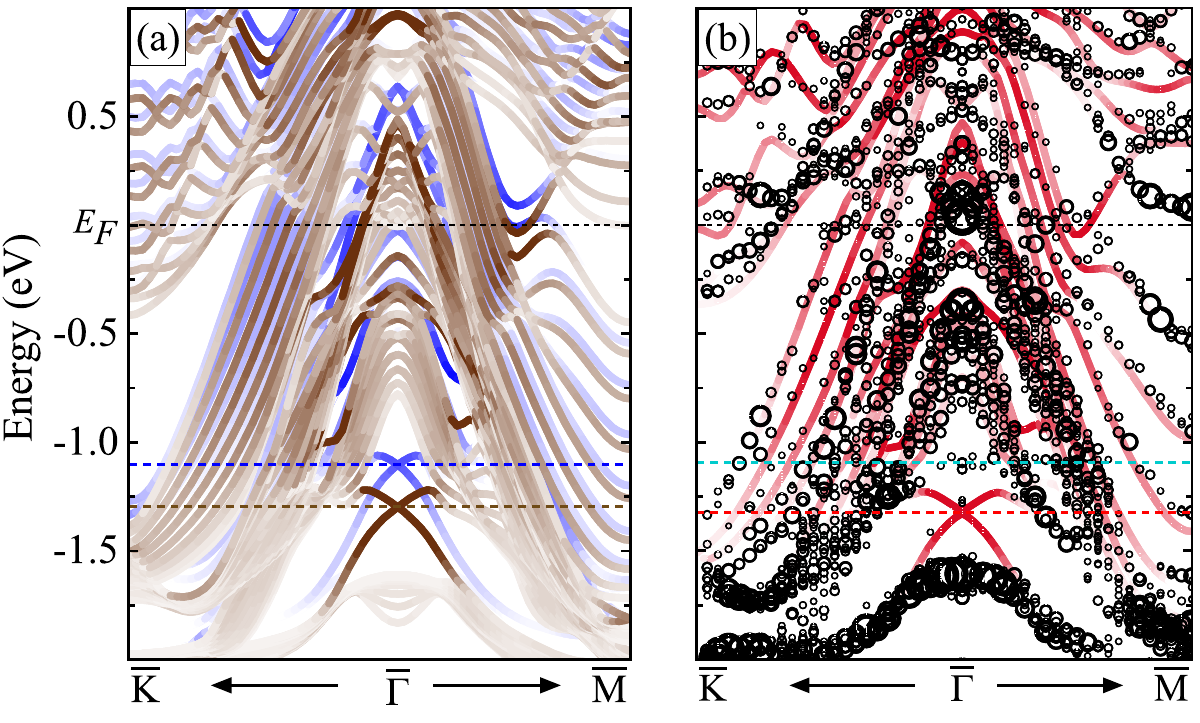}}
	\caption{ (a) DFT calculated surface band structure with 6\% Te vacancy at the surface incorporated using  VCA method (blue line) overlaid on pristine surface band structure for 1$\times$1$\times$10 slab configuration (brown line). (b) unfolded effective band structure using supercell approach with 11\% Te vacancy on the surface (black symbol), overlaid on pristine surface band structure for 1$\times$1$\times$5 slab configuration (red line).} \label{FigS6}
\end{figure}

\begin{figure*}[t]
	\centering
	{\includegraphics[width=0.95\textwidth]{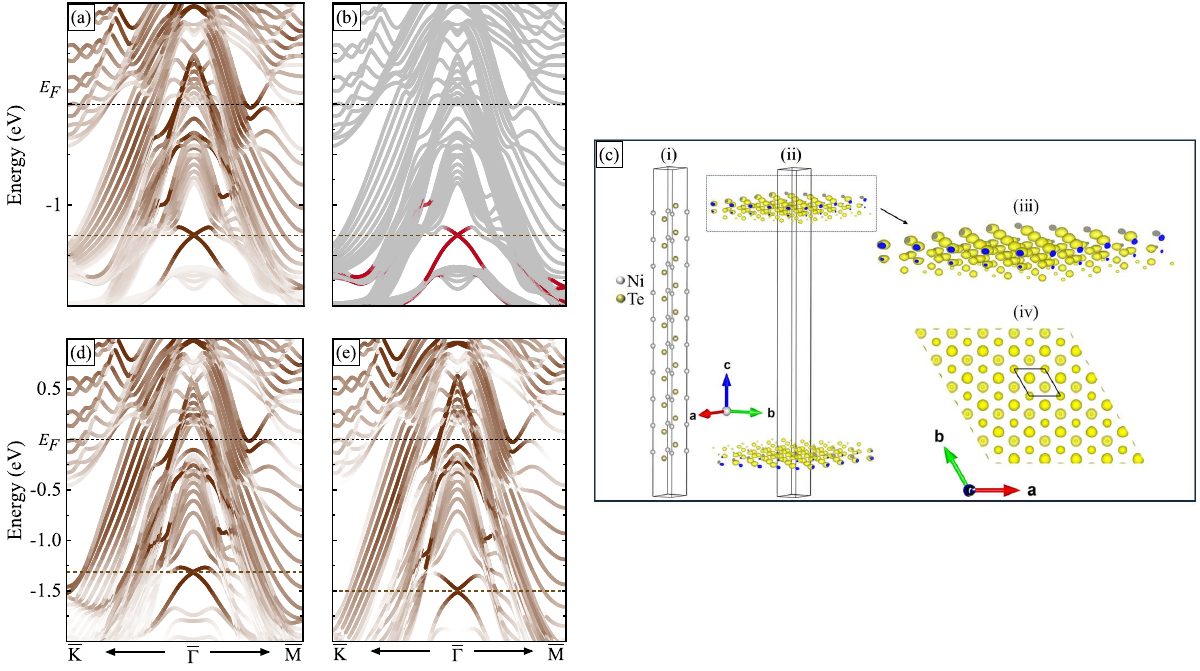}}
	\caption { DFT-calculated surface band structure highlighting the atomic contribution from (a) Te (brown) and (b) Ni atoms (red) at the surface. (c) Wave function (probability) plot corresponding to surface state SS2. DFT+$U$ calculated surface band structure for a  $U$ = 5 eV applied to 3$d$ states of (d) only bulk and (e) all Ni atoms.} \label{FigS7}
\end{figure*}

\section{ Effect of vacancy }

Transition metal dichalcogenides (TMDCs) are known to host a variety of defects formed during sample growth \cite{doi:10.1021/acsnano.0c09666}, regardless of the choice of synthesis method; among them, chalcogen vacancies are considered to be the most common defect \cite{PhysRevMaterials.6.084002,KANOUN2021101442}. Elemental analysis using energy dispersive X-ray analysis (EDX) for studied sample showed $\sim$1.5 \%  Te vacancy in the bulk crystal \cite{Mandal_2021}. The cleaved surface is most susceptible for chalcogen vacancies as seen in a variety of TMDCs \cite{https://doi.org/10.1002/adfm.201906556,nano14050410}.
Controlled defect engineering in TMDCs can also provide valuable opportunities to tailor the corresponding electronic properties \cite{PhysRevB.92.235408,Zhussupbekov2021}. To study the effect of chalcogen vacancy at the surface on the electronic properties of NiTe$_{2}$, we show the calculated surface band structure with 6 \% Te  vacancy at the surface  employing virtual crystal approximation (VCA) method \cite{PhysRevB.61.7877} within the 1$\times$1$\times$10 slab configuration in Fig. \ref{FigS6}(a). The band structure of pristine  surface and surface with Te vacancy has been shown by brown and blue color respectively. The bulk bands (lighter intensity color) do not show significant change in energy positions while all the surface states (higher intensity color) exhibit a significant shift towards $E_F$ for surface with Te vacancy. The energy position of the Dirac-like crossing formed by the surface state SS2, with and without vacancy, is indicated by the blue and brown dashed lines, respectively, exhibiting a substantial shift of $\sim$200 meV. To further verify the surface Te vacancy induced effects, we modeled the vacancy using supercell approach. However, to model a small \% of surface Te vacancy one needs a very large supercell making it computationally very expensive. Thus, we modeled a $\sim 11$ \% vacancy of Te atoms at the surface using Te-terminated 3$\times$3$\times$1 supercell of bulk NiTe$_{2}$ with 5 unit cell thick slab with 10 \AA\ of vacuum and removed one Te atom (out of 9) each from the top as well as bottom surfaces preserving inversion symmetry. The effective band structure after unfolding has been shown in Fig. \ref{FigS6}(b) using black symbols. For a direct comparison, we also show the band structure of pristine Te-terminated surface (1$\times$1$\times$5) using red color. The band structure of surface with Te vacancy matches very well with that of the pristine surface, except for a shift of all the bands related surface states towards $E_F$ suggesting that the vacancy induced effects are very similar to that observed using the VCA method. Surface state at the $E_F$ moves slightly above the $E_F$, while the energy position of the Dirac crossing shows a significant shift of 230 meV towards $E_F$ as shown by the red and cyan dotted lines.

 \begin{figure*}[t]
	\centering
	{\includegraphics[width=0.95\textwidth]{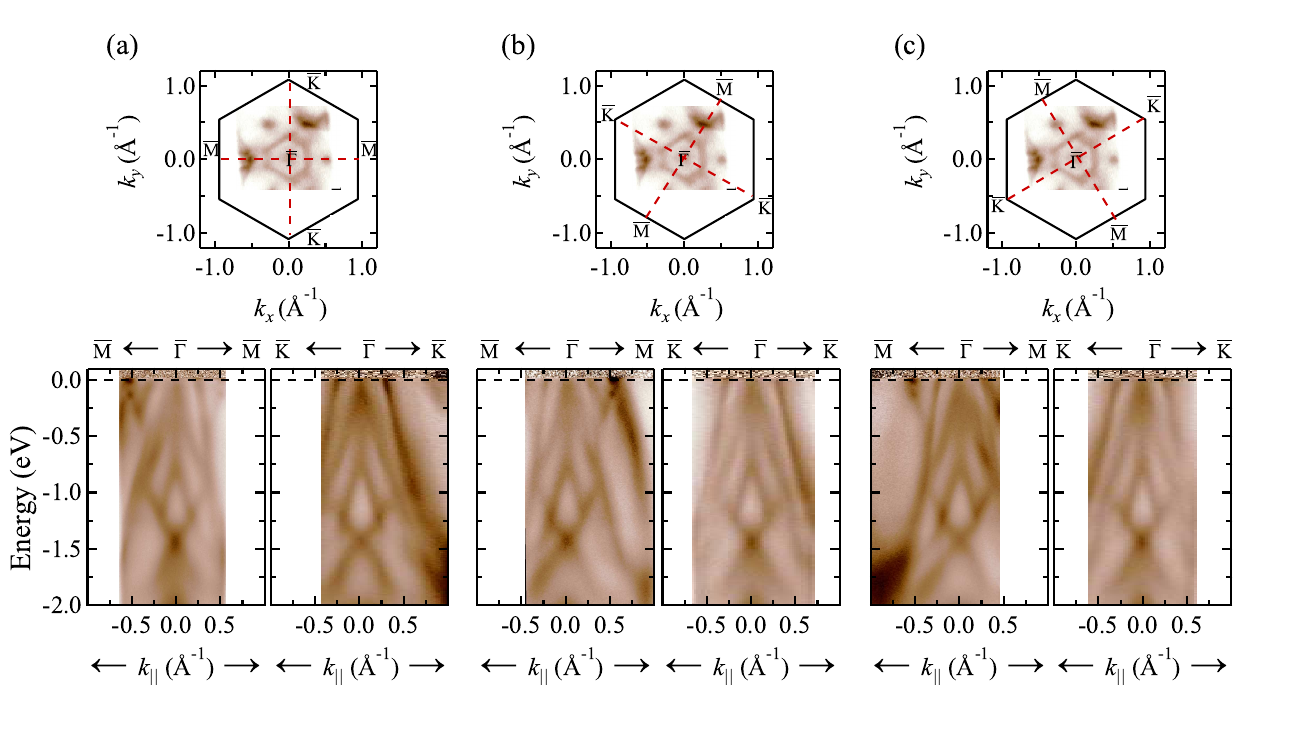}} 
	\caption{ (a)-(c) Experimental FS along with different momentum cuts (top panels) and corresponding band dispersions (bottom panels). Red dashed lines represent various momentum cuts on FS.} \label{FigS8}
\end{figure*} 
`

 \section{Calculated surface band structure and wave function }
We show the DFT calculated surface band structure for 1$\times$1$\times$10 slab configuration with Te and Ni atomic contribution from surface layer in Fig. \ref{FigS7}(a) and  Fig. \ref{FigS7}(b), respectively. As expected, all the surface states exhibit a dominant contribution from surface Te atoms, however, SS2 forming Dirac-like crossing exhibit strongly hybridized Ni $3d$ and Te 5$p$ character leading to a notable Ni contribution while it is negligible in  SS0 and SS0$^{\prime}$.  The real-space wavefunction (probability) corresponding to the topological surface state SS2 at $\Gamma$ has been shown in Fig. \ref{FigS7}(c). The different panels (i), (ii), (iii) and (iv), show the 10-unit cell thick slab with atoms, wavefunction plot without atoms and enlarged view of the wavefunction in different orientations respectively. The figure confirms that SS2 is primarily localized at the surface layer, with contributions from Te atoms as well as Ni atoms. This indicates significant hybridization between Te-$p$ and Ni-$d$ orbitals, forming SS2 which leads to strongly correlated nature of SS2. Strong correlation among $3d$ electrons of surface Ni atoms could significantly influence SS2 and it is important to include correlation effects  within DFT for an accurate and comprehensive description of various surface states.
To investigate the impact of electron correlation in the bulk and the surface electronic structure, we show the results of band structure calculation performed within DFT+$U$ method ($U$ = 5 eV for Ni 3$d$ states) with $U$ applied to only bulk Ni atoms (except for Ni in the top and bottom slabs) and to all Ni atoms in Fig. \ref{FigS7}(d) and Fig. \ref{FigS7}(e) respectively. The energy position of the surface state SS2 is shown by brown dashed line. Notably, the inclusion of electron correlation in the bulk only has minimal influence on all the surface states. However, as previously discussed (also in the main text), correlation induced bulk Ni-derived states shift towards lower energy in contrast to the experimental results. Interestingly, inclusion of electron correlation to all the Ni atoms leads to shift of the surface state SS2 towards a lower energy and appears at $-$1.50 eV. These results suggest that the surface is strongly correlated than the bulk leading to modification in the energy position of Dirac-like conical crossing of the topological surface state in NiTe$_2$.

\section{ ARPES spectra }
 ARPES FS, obtained using He {\scriptsize I} radiation corresponding to $k_z$ = 0.34 $c^*$, exhibit a three-fold symmetry (DFT calculated bulk FS in Fig. \ref{FigS3}(g) and surface FS in Fig. \ref{FigS3}(j)). Here, for completeness, we show the band dispersion (in the bottom panels) along three different momentum cuts (in the top panels) as shown  in the Fig. \ref{FigS8}(a)-(c). Band dispersions are very similar along all $\overline{\Gamma}$-$\overline{\textrm{K}}$ directions. However, due to bulk states forming gear shaped electron pocket along alternate $\overline{\Gamma}$-$\overline{\textrm{M}}$ direction, makes the FS three-fold symmetric with respect to band dispersions obtained along $\overline{\Gamma}$-$\overline{\textrm{M}}$ directions. The isolated SS0 \& SS0$^{\prime}$ surface states (without nearby bulk band) are also clearly evident along $\overline{\Gamma}$-$\overline{\textrm{M}}$ direction in the bottom panel of Fig. \ref{FigS8}(c).